\documentclass{emulateapj}

\usepackage{xcolor}
\usepackage{epsfig}
\usepackage{natbib}
\usepackage{slashbox}

\shorttitle{Angular Momentum Transport in Hot Jupiters}

\shortauthors{C. Yu}

\begin{document}


\title{ Thermally Driven  Angular Momentum Transport in Hot Jupiters}

\author{Cong ~Yu\altaffilmark{1,2}}
\altaffiltext{1}{School of Physics and Astronomy, Sun Yat-sen University, Zhuhai,
519082, China; {\tt yucong@mail.sysu.edu.cn}}

\altaffiltext{2}{Lunar and Planetary Science Laboratory, Macau University of Science and Technology, Macau, People’s Republic of China}

\begin{abstract}
We study the angular momentum transport inside the hot Jupiters under the the influences of gravitational and thermal forcing. Due to the strong stellar irradiation, radiative region develops on top of the convective region. Internal gravity waves are launched at the radiative-convective boundaries (RCBs). 
The thermal response is dynamical and plays an important role in the angular momentum transport.
By separating the gravitational and thermal forcing terms, we identify the thermal effects for increasing the angular momentum transport. 
For the low frequency (in the co-rotating frame with planets) prograde (retrograde) tidal frequency, the angular momentum flux is positive (negative). 
The tidal interactions tends to drive the planet to the synchronous state. 
We find that  the angular momentum transport associated with the internal gravity wave is very sensitive to relative position between the RCB and the penetration depth of the thermal forcing. 
If the RCB is in the vicinity of the thermal forcing penetration depth, even with small amplitude thermal forcing, the thermally driven angular momentum flux could be much larger than the flux induced by gravitational forcing. The thermally enhanced torque could
drive the planet to the synchronous state in as short as a few $10^4$ years. 
 

\end{abstract}




\section{INTRODUCTION}

Some short period exoplanets are close to the host stars. 
The strong tidal responses may take place and lead to the 
evolution of the spin and orbital properties of the planets. 
It is well established that tides play an essential role in shaping the 
dynamical architecture of planetary system (Ogilvie 2014, for a review). 
When the exoplanets undergo non-synchronous rotation, the periodic variation
of the stellar irradiation give rise to thermal tides. 
Gold \& Soter (1969, hereafter GS) originally proposed the concept
of thermal tide torques to explain Venus' asynchronous spin rate. 
Arras \& Socrates (2010) stressed the role of thermal tides for the 
hot Jupiter. They found that thermal tides could induce large asynchronous
spin, and generate tidal heating rates more than sufficient to power
the observed radii. 
Goodman (2009a) pointed out the GS ansatz does not faithfully represent the fluid motion 
induced by time-dependent heating in a completely fluid atmosphere. Gu \& Ogilvie (2009) studied the diurnal tides in a non-synchronized hot Jupiters and found that the retrograde thermal forcing generates negative angular momentum flux. 
Recently they revisited this problem with two-stream atmosphere treatment and found that
the self-absorption plays an important role counteracting the radiative damping (Gu et al. 2019). The thermal tides in rotating hot Jupiters are also explored (Auclair-Desrotour \& Leconte 2018; Lee \& Murakami 2019). 

Hot Jupiters are Jupiter-mass planets near their central host (within $\sim$0.1 au). 
They are subject to the strong irradiation and a deep radiative outer layer appears 
on the top of the convective interior (Guillot 2005). Tidal effects are also important
for their internal structure and evolution. In the radiative region with positive 
entropy gradient, internal gravity waves can propagate in these stably 
stratified regions. In stratified atmosphere, gravity waves are readily generated by many mechanisms,
thermal and gravitational. The momentum and energy transported by the internal 
gravity wave needs to be addressed. Zahn (1977) considered the tidal forcing of low-frequency internal gravity waves in early-type stars, which have convective cores and radiative envelopes. The structure of early-type stars is similar to hot Jupiters.

The angular momentum transport within the planet is important in two aspects. 
First, the spin of the planet is directly related to the angular momentum transport. 
Lubow et al. (1997) applied Zahn’s dynamical tide to hot Jupiters, short-period extrasolar giant planets in which stellar irradiation makes the atmosphere stably stratified to 
a depth on the order of 100 bars.
They predicted that 51 Peg B could be synchronized in about $\sim 10^5$ years. 
Second, it is closely related to the atmosphere circulation. 
Both observations and numerical simulations have been performed to reveal that
the planet atmospheres possess differential banded structures and vertical shear. How the angular momentum is transported and redistributed within the planet is of vital importance 
to understand the atmosphere flow structure (Cho 2008).  Tidally-induced angular momentum transport have also been widely studied for early-type star (Zahn 1975, 1977; Goldreich \& Nicholson 1989), solar type stars (Goodman \& Dickson 1998; Ogilvie \& Lin 2007), as well as white dwarfs (Fuller \& Lai 2012).

This paper is structured as follows: in \S 2 we introduce the equilibrium structure of hot Jupiter.
In \S 3 we will discuss basic dynamical equations. The angular momentum transport
within the hot Jupiter will be discussed  in \S 4. 
Conclusions and discussions are given in \S 5.



\section{Equilibrium Structure of Hot Jupiter}
The equilibrium structure of the strongly irradiated planet reads 
\begin{equation}
\frac{d P}{d r} = - \rho g \ ,
\end{equation}
\begin{equation}
\frac{d M_r}{d r} = 4 \pi \rho r^2 \ ,
\end{equation}
where $g = {G M_r}/{r^2}$ is the gravitational acceleration, $G$ is the gravitational constant, $P$ is the pressure, $\rho$ is the
density, and $M_r$ is the mass enclosed inside the radius $r$.
The planet mass is taken as $m_p = 1.7 \times 10^{30}$g = 0.89$M_J$ and the radius is $r_p = 9\times 10^{9}$cm = 1.28$R_J$.  
The equation of state $\rho = \rho(P)$ is based on Arras \& Socrates (2010),
\begin{equation}
\rho = e^{-P/p_b}\left(  \frac{P}{a^2} \right) + \left( 1 - e^{-P/p_b} \right) \sqrt{\frac{P}{K_c}}  \ ,
\end{equation}
where $K_c = G R_J^2$, $a^2 = \sqrt{K_c p_b}$, and $p_b$ is the pressure at the base of the radiative region and we choose $p_b = 100$ bar. The planet is composed of a
nearly isentropic convective core and a thin radiative envelope. Since convective core is neutrally stable, we choose $\Gamma_1 = 2$ and the Brunt-Vasala frequency in the core is approximately zero.  
We note that this model, though simple, well reproduces the main feature of 
hot Jupiters. The density at the bottom of the radiative region is roughly 1\% of the mean
density. The thickness of the radiative region covers about 2\% of the radius of the hot Jupiter.


\section{Basic Dynamical equations}
We focus on the semi-diurnal components. The linear displacement, pressure and density perturbations are all proportional to $Y_{22}(\theta, \phi) e^{-i \sigma t}$, where $\sigma$ = $2 (n-\Omega_s)$. $n$ is the orbital frequency and $\Omega_s$ is the spin frequency of the hot Jupiter. The external tidal potential is $U$ = $- \sqrt{3\pi/10} \ n^2 r^2$. 

The zero-frequency equilibrium gravitational tides,  which denote the hydrostatic response of the planet to the tidal potential and can be separated from the dynamical response. The idea of equilibrium tides breaks down in regions that are neutrally 
stratified and therefore, the forcing frequency cannot be safely set to zero 
(Arras \& Socrates 2010). In other words, the hydrodynamic response in 
neutrally stratified regions is inherently a non-equilibrium tide. As a result, we decompose the planet response  as dynamical and equilibrium part  
$\xi_r = \xi_r^{(d)} + \xi_r^{(e)}$, $\delta P = \delta P^{(d)} + \delta P^{(e)}$, and $\delta \rho = \delta \rho^{(d)} + \delta \rho^{(e)}$.
The equilibrium part can be written as (Goldreich \& Nicholson 1989) 
\begin{equation}
\delta P_{\rm gr}^{(e)} = - \rho U \ , 
\end{equation}
\begin{equation}
\xi_{r, \rm gr}^{(e)} = - \frac{U}{g} \ , 
\end{equation}
\begin{equation}
l(l+1) r \xi^{(e)}_{\perp, \rm gr} = - \frac{d}{dr}\left( \frac{r^2 U}{g} \right) \ .
\end{equation}
The zero frequency equilibrium tides only exist for gravitational tides. We use the subscript `gr' to denote the equilibrium tides caused by the tidal gravitational potential. Note that we can not separate thermal equilibrium tides. As a result, we include all the thermal response in the dynamical components.
We adopt $y_1 = r^2 \xi_r^{(d)}$, $y_2 = {\delta P^{(d)}}/{\rho}$,   and $y_3 = {\delta \rho^{(d)}}/{\rho}$
as our basic variables and the dynamical equations for the linear response 
are as follows (Unno et al. 1989; Goodman \& Dickson 1998),
\begin{equation}
\frac{d y_1}{d r} = A_{11} y_1 + A_{12} y_2 + A_{13} y_3 - {l(l+1)} r \xi_{\perp, \rm gr}^{(e)} \ ,
\end{equation}
\begin{equation}
\frac{d y_2}{d r} =   A_{21} y_1 + A_{22}  y_2 + A_{23} y_3 + \sigma^2 \xi_{r, \rm gr}^e \ ,
\end{equation}
where 
\begin{equation}
A_{11} =  - \frac{d \ln \rho}{dr} \ , A_{12} = \frac{l(l+1)}{\sigma^2} \ , 
A_{13} = - r^2 \ ,
\end{equation}
\begin{equation}
A_{21} = \frac{\sigma^2}{r^2} \ , A_{22} = A_{11} \ ,  A_{23} = - g \ .
\end{equation}

In contrast to the gravitational response, it is not possible to find the equilibrium thermal response in the zero frequency limit. 
That means the thermal response is intrinsically a time-dependent dynamical process. 
The thermal response is included in the energy equations, which can be simply written as 
(note that $y_1$, $y_2$, $y_3$ denote the dynamical responses)%
\begin{equation}
\frac{N^2}{g r^2} y_1 + \frac{1}{c_s^2} y_2 - y_3 = \frac{\Delta s}{c_p} \ ,
\end{equation}
where $c_s^2 = \Gamma_1 P/\rho$ is the adiabatic sound speed and $\Delta s$ is the Lagrangian entropy perturbation. The Brunt-Vaisala frequency and the sound speed are related as 
\begin{equation}
- \frac{d \ln \rho}{d r} = \frac{N^2}{g} +\frac{g}{c_s^2} \ .
\end{equation}

The thermal forcing can be cast as (Arras \& Socrates 2010; Auclair-Desrotour \& Leconte 2018; Lee \& Murakami 2019)  
\begin{equation}
\frac{\Delta s}{c_p} = \frac{i}{\sigma \tau_{{\rm th}}} e^{- p/p_{\rm depth}} \ , 
\end{equation}
where $p_{\rm depth}$ is the location that the external heating can penetrate. The deepest location that the external heating can reach is the base of the radiative region. From this equation, we can see that the zero frequency limit $\sigma \rightarrow 0$ is not appropriate for the non-adiabatic calculations.
The perturbed heating due to self-absorption of thermal emissions can be significant against Newtonian damping (Gu et al. 2019). 
Note that thermal forcing may even come from the heat from the rocky core if the envelope and the rocky core are experiencing differential rotations (Ginzburg et al. 2016). 

The thermal tides are dynamical and we can not find thermal equilibrium tide. We find that in the radiative region, the term $\frac{N^2}{g} \xi_r^{(d)}$ rapidly oscillates with increase amplitude towards the surface. 
The amplitude of this term is much larger than $\frac{\Delta s}{c_p}$.  Actually, this term is nearly cancelled by the $\frac{\delta \rho^{(d)}}{\rho}$. This is quite different from the equation (38) in Arras \& Socrates (2010), who proposed that the flow in radiative region compensates the thermal forcing and makes the quadrupole density variation vanish in the zero frequency limit.
In our calculation, the contrary is actually the case. It is the flow in the radiative region that dictates the quadrupole density variation in the low frequency limit. In the evanescent convective region, the term $\frac{1}{c_s^2} \frac{\delta P^{(d)}}{\rho}$ decays towards the center. It is nearly cancelled by $\frac{\delta \rho^{(d)}}{\rho}$.

Note that Equations (9),  (10), (11) can be combined together as a single second order non-homogenous ordinary differential equation (ODE). 
\begin{equation}
\frac{d^2 y_2}{ d r^2} + A \frac{d y_2}{d r} + B y_2 = C \ , 
\end{equation}
where 
\[
A = \left( \ln \frac{r^2\rho}{W}\right)^{\prime} \ , 
\]
\[
B = - \left[ \left( \frac{N^2}{g} \right)^{\prime} + \frac{N^2}{g} \left( \frac{r^2}{W}\right)^{\prime} - \frac{\sigma^2}{c_s^2} - \frac{W l (l+1)}{r^2 \sigma^2}\right] \ , 
\]
\[
C = \left( \sigma^2 \xi_{r, \rm { gr}}^{(e)} + g \frac{\Delta s}{c_p} \right)^{\prime} + \left( \ln \frac{r^2}{W}\right)^{\prime}\left( \sigma^2 \xi_{r, \rm { gr}}^{(e)} + g \frac{\Delta s}{c_p} \right)
\]
\[
- \frac{g}{c_s^2} \left( \sigma^2 \xi_{r, \rm { gr}}^{(e)} + g \frac{\Delta s}{c_p} \right) 
- \frac{W}{r^2} \left( r^2 \frac{\Delta s}{c_p } - l (l+1) r \xi_{\perp, \rm { gr}}^{(e)} \right) \ , 
\]
and $W=N^2 - \sigma^2$, $^{\prime} \equiv \frac{d}{dr}$. 

The source terms on the right hand side this ODE includes all the forcings. The terms related to $\xi_r^{(e)}$ and $\xi_{\perp}^{(e)}$ pertains to the gravitational forcing. Note that if the thermal forcing is neglected, we recover the equations in Savonije \& Papaloizou (1984). Those terms related to $\frac{\Delta s}{c_p}$ are the thermal forcing. The wave excitation mechanisms can be well studied with this single ODE. 

Since the radiative region is very thin (thickness  $\sim 2 \%$ of the planet radius). The internal gravity waves can be 
better resolved in the radiative region with the pressure 
as the independent variable (Watkins \& Cho 2009).
We adopt the relaxation technique to solve the two point boundary value problem (Press et al. 1992). We use staggered mesh techniques to take care of the algebraic equation (11). 


\subsection{Boundary Conditions}
The equations (6), (7), and (8) should be complemented with appropriate boundary conditions. 
At the inner boundary, we use the standard regularity condition requiring all variables to be finite (Unno et al. 1989)
\begin{equation}
r^2 \left(\xi^{(d)}_r + \xi^{(e)}_r \right) = \frac{2 r}{\sigma^2} \left( \frac{\delta P^{(d)}}{\rho}  \right)  \  . \
\end{equation}

Note that we are particularly interested in the regime that the driving frequency, $\sigma^2$, is much smaller than
the Brunt-Vaisala frequency, $N^2$, in the radiative region. In this regime, the internal gravity can propagate in the radiative region (i.e., not evanescent). 
Since the density of the planet envelope decreases with the radius, 
the amplitudes of the internal gravity waves
become larger as they propagate towards the planet surface. Nonlinear damping in the outer layer of the planet
would prevent the formation of discrete modes (Fuller \& Lai 2012). 

We assume that the waves are efficiently damped in the planet envelope and we implement the outgoing wave boundary conditions 
at the planet surface. 
For internal gravity waves, the vertical component of the group velocity 
and the phase velocity are in the opposite direction. 
For sub-synchronous planets, $\sigma = 2(n-\Omega_s) > 0 $,
the real part of the waver number for the outgoing waves should be  ${\rm Re}\{k_r\} < 0$ since the vertical component of group velocity is positive.
For super-synchronous planets, $\sigma = 2(n-\Omega_s) < 0 $, we select ${\rm Re}\{k_r\}>0$. Note that we implicitly include rotation for the the driving frequency $\sigma$ as did in Goldreich \& Nicholson (1989).

\section{Angular Momentum Transport}
The internal gravity waves are launched in a narrow region near the base of the radiative envelope, where the Brunt-Vaisala frequency matches the tidal frequency and the radial wavelength is long enough to connect to the tidal forcing.
As the internal gravity wave is launched and propagates in the planet, it carries an angular
momentum flux to the outer layers. The time averaged angular momentum flux reads
\begin{equation}
\dot{J} = 2 m \sigma^2 \rho r^3 {\rm Re} \left\{i \overline{\xi_r^{(d)}} \xi_{\perp}^{(d)} \right\}  
\end{equation}
where $\overline{\xi_r^{(d)}}$ represents the complex conjugate of ${\xi_r^{(d)}}$.
With outflow boundary conditions, the angular momentum can be negative or positive, which depends on the planet is prograde or retrograde. 


\subsection{Comparison with prior results }
In the convective core, the gravity waves are evanescent. They can propagate in the outer stably stratified radiative region. The RCB is a turning point where the WKB wavelength of gravity waves is infinite (Goldreich \& Nicholson 1989).  
Since the orbital period is long compared to the dynamical time of the star, the 
g-mode has a short radial wavelength, therefore, significant coupling between mode
and the tidal potential occurs only at the boundary between radiative and convective 
zone, where the wavelength of g-mode becomes large.
We have confirmed that our numerical results are consistent with the WKB relation
\begin{equation}
\xi^{(d)}_{\perp} \approx i \frac{k_r r}{l(l+1)} \xi_r^{(d)} \ .
\end{equation} 
When the thermal timescale is long enough, the planet response is nearly adiabatic. 
For the calculation with infinite thermal timescale, we are supposed to recover prior results.  Lubow et al. (1997) gave an empirical estimation about the angular momentum flux inside hot Jupiters, which reads
\begin{equation}
\dot{J}_{\rm empirical} \sim \rho_b r_p^5 \sigma^2 \left( \frac{H}{r_p}\right)^{2/3} \left(\frac{\sigma}{n} \right)^{5/3} \left(\frac{n}{\omega_p}\right)^{14/3} \ ,
\end{equation}
where  $\omega_p = (G m_p/r_p^3)^{1/2}$,  $\rho_b$ is the density, $H$ is the scale height at the base of the radiative region.  In our calculation, $\rho_b \sim  3 \times 10^{-3} $, $H \sim 2 \times 10^{-3} r_p$, the orbital frequency $n = 8.9\times 10^{-2}$. Note that the numerical value of the frequency, density and scale height are measured in  $\omega_p$, $\rho_{\rm mean} = m_p/r_p^3$, and $r_p$, respectively.

\begin{figure}
\includegraphics[scale=0.53]{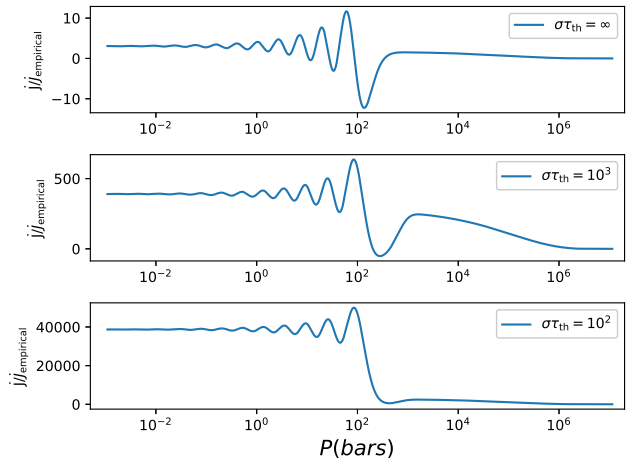}
\caption{
The variation of angular momentum flux with the radius. The thermal penetration depth is $p_{\rm depth} = 10^8$ dyne/cm$^2$. The gravitational potential corresponds to $n = 8.9 \times 10^{-2}$, which is equivalent to an orbital period about $\sim 2$ days. The driving frequency is $\sigma = 10^{-2}$. Note that the frequency is measured in  $\omega_p = \left(G m_p/r_p^3 \right)^{1/2}$. The angular momentum flux $\dot{J}$ is measured  in $G m_p^2/r_p$.
}
\label{fig1}
\end{figure}
In the upper panel of Figure 1, we show the angular momentum flux without the thermal forcing. Note that near the surface of the planet, the angular momentum flux is nearly a constant\footnote{This is consistent with the equation,
${ \rho r^2 ( \xi_{\perp}^{(d)} )^2}/{k_r} = \rm constant$, in Goldreich \& Nicholson (1989). }.
This is the net angular momentum flux that the internal gravity waves carry. 
We find that the order of magnitude of the angular momentum flux are consistent with the empirical estimation, though the exact numbers may be different by a factor of a few.

The variation of angular momentum flux for $\sigma = 10^{-2}$, $2\times 10^{-2}$ without thermal forcing are shown in the upper panel of Fig. 1 and Fig. 2, respectively. 
It is clear from Equation (20) that the angular momentum flux becomes smaller for lower driving frequency ($\dot{J} \propto \sigma^{11/3}$), the lower the driving frequency, the longer the synchronization timescale\footnote{Note that this is different from the scaling $\dot{J} \propto \sigma^2$ in Arras \& Socrates (2010), which is obtained in terms of the zero frequency equilibrium tides.}.
Our numerical results are consistent with the theoretical estimation. When the thermal forcing is ignored, the synchronization timescale for the case of low driving frequency $\sigma = 10^{-2}$,  is $\sim 50$ Myrs. For $\sigma = 2\times 10^{-2}$, the synchronization timescale is $ \sim 2$ Myr. This means that, when a sub-synchronous planet is approaching the synchronous state, the spin-down process gradually becomes slower.  

\subsection{Angular Momentum Flux Enhancement } 
It is clear that, from Equation (13), when thermal forcing is considered, new terms associated with the thermal forcing appears on the right hand side of the wave equations. 

\begin{figure}
\includegraphics[scale=0.53]{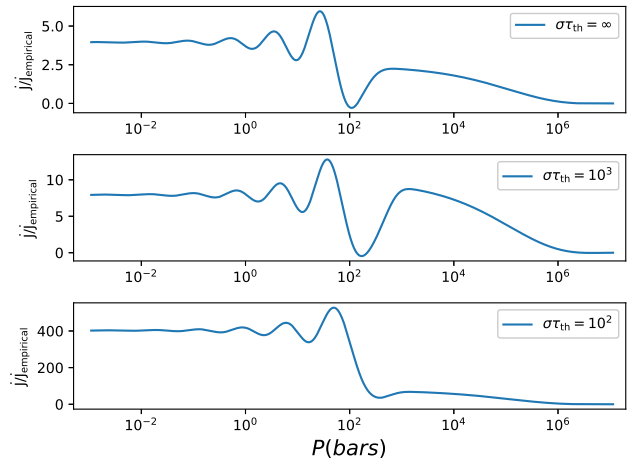}
\caption{
The driving frequency is $\sigma = 2 \times 10^{-2}$.  The angular momentum flux can be greatly enhanced by thermal forcing if the thermal timescale is sufficiently short. Note that the frequency is measured in  $\omega_p = \left(G m_p/r_p^3 \right)^{1/2}$. The angular momentum flux $\dot{J}$ is measured  in $G m_p^2/r_p$.
}
\label{fig2}
\end{figure}

To understand how the driving frequency affects the enhancement,  we perform calculations with different driving frequencies. 
The variation of  the angular momentum flux for frequencies, $\sigma =10^{-2}$ and $\sigma =  2\times  10^{-2}$ are shown in Fig. 1 and Fig. 2, respectively. 
We can compare the angular momentum flux without thermal forcing and those with various thermal forcing strengths from these two figures.



The upper panels show the result without thermal forcing. The middle and lower panels show 
the angular momentum flux distribution with different thermal forcing strengths. With a low driving frequency $\sigma = 10^{-2}$,  the ratio between the net angular momentum flux with and without thermal forcing is approximately $10^2$ for $\sigma \tau_{\rm th} = 10^3$. The ratio increases 
to $\sim 10^3$ for $\sigma \tau_{\rm th} = 10^2$. It is obvious that  the net angular momentum flux become higher when the thermal forcing becomes stronger.

For a higher driving frequency $\sigma = 2\times 10^{-2}$, the net angular momentum flux increases by a factor of 2 when $\sigma \tau_{\rm th} = 10^3$. For stronger thermal forcing with $\sigma \tau_{\rm th} = 10^2$, the net angular momentum flux increases by a factor of $10^2$. 
Our calculation also shows that the enhancement of angular momentum transport is more pronounced for low driving frequency. When the thermal forcing is considered, the timescale for synchronization can be as short as $\sim 10^4$ years. 

To be more realistic, we have performed non-adiabatic calculation based on the Newtonian cooling (Auclair-Desrotour \& Leconte 2018). The corresponding energy equation becomes
\begin{equation}
\frac{N^2}{g} \xi_r + \left( 1 + { i \frac{\Gamma_1}{\sigma \tau_{\rm th}} }\right) \frac{\delta P}{\rho c_s^2} +  \left( 1 + { i \frac{1}{\sigma \tau_{\rm th}} }\right) \frac{\delta \rho}{\rho} 
= i\frac{e^{-p/p_{\rm depth}}}{\sigma \tau_{\rm th}} \ . 
\end{equation} 
When $\sigma \tau_{\rm th} \gg1$, the radiative damping plays a minor role in the transport of angular momentum. For lower value of $\sigma \tau_{\rm th}$, the radiative damping becomes important.   
For $\sigma \tau_{\rm th} \sim 10$, if we do not consider radiative damping effects, the enhancement could be even higher than the case of $\sigma \tau_{\rm th} = 10^2$. However, when the radiative damping is included, the enhancement could only reach a factor of a few $10^3$. In the regime of $\sigma \tau_{\rm th} \sim 1$, the damping is so strong that we can not observe the thermal enhancement. It is clear that the thermal enhancement depends on the the parameter $\sigma \tau_{\rm th}$ in a non-monotonic way.  When $\sigma \tau_{\rm th}  > 10^3$, the excitation becomes weak, the thermal enhancement is not obvious. In the range between the two ends, i.e., $\sigma \tau_{\rm th} \sim 10-10^3$, the thermal enhancement could be maximized. It is worthwhile to note that, in the low $\sigma \tau_{\rm th}$ regime, more rigorous treatment of the radiative transfer is necessary to understand the non-adiabatic effects on the thermal tides (Gu et al. 2019).

In addition, we also perform calculations for  super-synchronous frequencies ($\sigma < 0 $). We find the net angular momentum flux can be negative for the super-synchronous planet. The thermal enhancement of the angular momentum transport still operates for super-synchronous planets. That means the super-synchronous planet experience an accelerated  spinning down process due to the enhanced negative torque.

For the low driving frequency, the dynamical tidal components, both gravitational and thermal,  behave in a quite similar manner. That is to say that the asymptotic behavior (low frequency limit, or short wavelength limit) of dynamical gravitational tides in Goldreich \& Nicholson (1989) still applies for dynamical thermal tides. 
According to Equation (16) and (17), we have
\begin{equation}
\dot{J} \sim {\rm Re}\left\{i \overline{\xi_r^{(d)} } \xi_{\perp}^{(d)}\right\} = {\rm Re} \{ - k_r \} \frac{r \left|\xi_r^{(d)}\right|^2}{l(l+1)}  \  .
\end{equation}
Note that the absolute value of $\left| \xi_r^{(d)} \right|^2$ is always positive. 
As a result, the sign of angular momentum flux is opposite to the sign of real part of the wavenumber in the radial direction. Since we adopt the outgoing boundary condition,  ${\rm Re}\{k_r\} < 0$ for sub-synchronous planets and  ${\rm Re}\{k_r\} > 0$ for super-synchronous planets. It is clear that the net angular momentum flux is positive for the sub-synchronous planets, and negative for super-synchronous planets. With higher tidal frequency, the above WKB approximation is no longer valid. The torque can be opposite to the above WKB analysis, generating oscillating-signed values of the torque (Arras \& Socrates 2010).


%

\subsection{Dependence on the relative position between the RCB and  the thermal forcing}
Note that the enhancement is not due to the resonance between the driving frequency and 
the planet eigen-frequency (Arras \& Socrates 2010). In our case, the relative location between RCB and thermal forcing and amplitude of the thermal driving is crucial for the enhancement. 
Hot Jupiters with deep RCB has been put forward to explain the 
radius anomaly (Youdin \& Mitchell 2010; Ginzburg \& Sari 2015). 
Recent studies show that the RCB can be shallower (Thorngren et al. 2019).  
The depth of the RCB of hot Jupiters is still an uncertain parameter, which is now under active investigations.
Here the treat the relative position of the RCB and the penetration depth as a free parameter. 
It would be interesting to see how relative position between the RCB and the penetration depth of the thermal forcing  changes the dynamical response of the planet. 

We show the angular momentum flux within the planet in Fig. 3,  with the thermal forcing depth $p_{\rm depth} = 10^7$ dyne cm$^{-2}$.  Note that in our equilibrium model, the RCB is at $10^8$ dyne cm$^{-2}$. In this case, the penetration depth is shallower than the RCB. The driving frequency is $\sigma = 10^{-2}$. 
We can clearly identify that the net angular momentum flux is almost the same for different thermal forcing strengths. 
If the penetration is shallower than the RCB, we find that the wave amplitude around the thermal forcing is amplified. However, the thermal forcing could hardly affect the net  angular momentum flux within the planet. When we compare Fig. 1 and Fig. 3, we can conclude that, if the penetration depth is in the vicinity of the RCB, the thermal enhancement can be maximized. 

\begin{figure}
\includegraphics[scale=0.53]{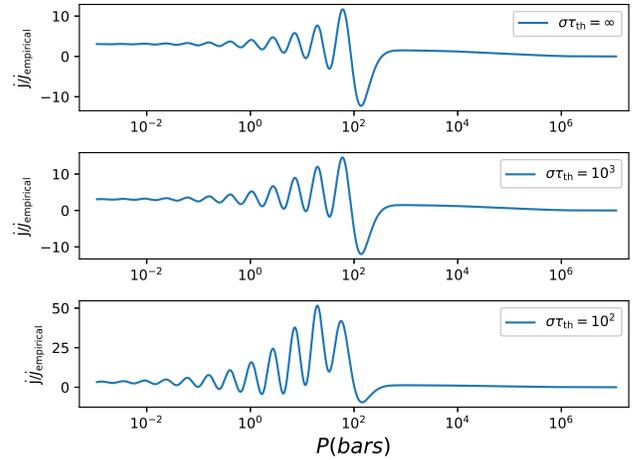}
\caption{
The variation of angular momentum flux with the radius. The thermal forcing penetration depth is $p_{\rm depth} = 10^7$ dyne/cm$^2$. We could identify that large amplitude in angular momentum flux is produced. But the net angular momentum flux at the surface is not greatly affected. The gravitational potential corresponds to $n = 8.9 \times 10^{-2}$, which is equivalent to an orbital period about $\sim 2$ days. The driving frequency is $\sigma = 10^{-2}$. In our calculation, the frequency is measured in  $\omega_p = (G m_p/r_p^3)^{1/2}$. The angular momentum flux, $\dot{J}$, is measured in $G m_p^2/r_p$.
}
\label{fig2}
\end{figure}

\section{Conclusions and Discussions}\label{sec:diss}

In this paper, we explore the transfer of angular momentum  inside the hot Jupiters under the the influences of gravitational and thermal forcing. We focus on the internal gravity waves which are excited at the radiative-convective boundaries (RCBs). By separating the equilibrium and dynamical responses, we make clear  the thermal effects for increasing the angular momentum transport. 

For the prograde hot Jupiters, the angular momentum flux is positive. For the retrograde hot Jupiters, the flux is negative. As a result, the tidal interactions tends to drive the planet to the synchronous state. Our calculation shows that the angular momentum transport  is very sensitive to the 
relative position of the thermal penetration depth and RCB.  If the penetration depth is much shallower than the RCB, large amplitude perturbation could be generated inside the radiative region, but the thermal forcing has minor effects on the net angular momentum flux. If the penetration depth is in the vicinity of RCB, even with small amplitude thermal forcing, the thermal forcing can significantly enhance the net angular momentum flux within the planet. 
The thermally enhanced torque could drive the planet to the synchronous state in as short as a few $10^4$ years.

In this work, we use a very simple planet interior model to study the tidal response. 
Since the tidal response depends critically on the interior structure of the planets, there are 
still some uncertainties concerning the behavior of low-frequency waves in planetary 
interiors. The RCB location is important for the tidal excitation of internal gravity waves. 
However, the exact location of RCB is still uncertain, which is worth further investigation (Ginzburg \& Sari 2015; Komadeck \& Youdin 20; Thorngren et al. 2019). In addition, how the multi-layered semi-convection structure affect the tidal responses of hot Jupiters would be an interesting issue to be further investigated. 

An efficient coupling between the radiative envelope and the deeper convective interior to is necessary to change the whole planet's spin in $\sim10^4$ years. The coupling is crucial for the synchronization process. As revealed in Goldreich \& Nicholson (1989), the spin-down is a outside-in process. In this case, the outer radiative region and inner convective region may experience differential rotation, the hydrodynamic instability associated with such velocity shear may produce an efficient coupling between these two regions. In addition, the magnetic field in hot Jupiter may also provide such couplings between these two regions. 

If the synchronization timescale is reduced, the tidal heating may operate in a short 
period of time, the radius anomaly may not well explained by the tidal heating mechanism. This work may also be important for the wind-flow interactions. 
Multi-layer convection could also alter the conclusion in this paper. It would be interesting
to further study how g-mode affect the angular momentum transport inside multi-layer
convection planet (Andre et al. 2019). 

Some short period Super-earths may also subject to the strong irradiation. 
The cooling history of super-Earths are closely related to location the radiative-convective boundary and turbulence may expand the radiative zone (Yu 2017). The RCB variations in super-Earth interior may also have important implication for the tidal interaction during the evolution of super-Earths. If the rocky core's heat is released periodically (Ginzburg et al. 2016; Liu et al. 2019), the heat perturbation would appear in the convective region, the core heat release process would affect the angular momentum transport within the super-Earths\footnote{The stellar irradiation could hardly reach the convective region, but the rocky core is embedded within the convective region. When we impose the thermal forcing inside the convective region, the angular momentum flux can also be greatly enhanced.}.   How the thermo-mechanical processes revealed in this paper affect the  evolution of the hot Jupiter needs to be further investigated.


Internal gravity waves are significant for the large scale flow and temperature structure in the exoplanetary atmosphere. Due to the strong irradiation, the acceleration and heating effects of gravity waves can be evident on hot Jupiters. The angular momentum transport by thermal effects may also have implications for the 
super-rotation of zonal jets, winds, clouds of hot Jupiters (Showman \& Polvani 2011; Zhang et al. 2015; Yang et al. 2019). The angular momentum
transport may generate vertical shear. Mean-flow and wave interactions may naturally occurs
in the hot Jupiter's atmosphere.

Rotation should be considered more carefully. Traditional approximation are widely
used in the literature(Lai 1997; Auclair et al. 2018). More rigorous treatment shows that the size
of the central rocky core plays an essential role (Ogilvie \& Lin 2004; Goodman 2009b; 
Lee \& Murakami 2019; Liu et al. 2019). How does rotation affects the g-mode's behavior addressed in this paper remains to be answered.

\acknowledgments
We thank the anonymous referee for the thoughtful comments that greatly improve this paper. This work is supported by Open Projects Funding of Lunar and Planetary Science Laboratory, MUST — Partner Laboratory of Key Laboratory of Lunar and Deep Space Exploration, CAS (Macau FDCT grant No. 119/2017/A3). C.Y. has been supported by the National Natural Science Foundation of China (grants 11373064, 11521303, 11733010, 11873103), Yunnan National Science Foundation (grant 2014HB048), and Yunnan Province (2017HC018). 







\begin{thebibliography}{99}

\bibitem[Andre et al]{Andre2019}
Andre, Q.,  Mathis, S., \& Barker, A. J., 2019, A\&A, 626, 82

\bibitem[Arras \& Socrates]{AS10}
Arras, P., Socrates, A., 2010, ApJ, 714, 1

\bibitem[Auclair-Desrotour \& Leconte ]{AL2018}
Auclair-Desrotour P., Leconte J., 2018, A\&A, 613, A45

\bibitem[Cho]{C08}
Cho J. Y. K., 2008, Phil. Trans. R. Soc. A., 366, 4477

\bibitem[Fuller \& Lai]{FL12}
Fuller, J.,  Lai, D., 2012, MNRAS, 421, 426

\bibitem[Ginzburg \& Sari]{GS15}
Ginzburg, S., Sari, R., ApJ, 2015, 803, 111

\bibitem[Ginzburg \& Sari]{GSS16}
Ginzburg, S., Schlichting H. E., \& Sari R.,  ApJ, 2016,  825, 29

\bibitem[Gold \& Stoer]{GS69}
Gold,T., Stoer S., 1969, Icarus, 11, 356

\bibitem[Goldreich \& Nicholson]{GN89}
Goldreich, P., Nicholson, P. D., 1989, ApJ, 342, 1079

\bibitem[Goodman \& Dickson]{Goodman98}
Goodman, J., Dickson, E., S., 1998, ApJ, 507, 938

\bibitem[Goodman]{Goodman09a}
Goodman J., 2009, arXiv:0901.3279

\bibitem[Goodman]{Goodman09b}
Goodman J., Lackner C., 2009, ApJ, 696, 2054

\bibitem[Gu \& Ogilvie]{GO09}
Gu P. G., Ogilvie, G. I., 2009, MNRAS, 395, 422

\bibitem[Gu et al.]{G19}
Gu P. G. Peng D. K., \& Yen C. C., 2019, ApJ, 887, 228


\bibitem[Guillot]{G05}
Guillot T., 2005, Annu. Rev. Earth Planet Sci., 33, 493

\bibitem[Komacek \& Youdin]{KY17}
Komacek, T. D., Youdin A. N., 2017, ApJ, 844, 94

\bibitem[Liu et al.]{Liu19}
Liu S. F., et al., 2019, Nature, 572, 355

\bibitem[Lubow et al.]{Lubow97}
Lubow, S., Tout, C. A., \& Livio, M., 1997, ApJ, 484, 866

\bibitem[Lai]{L97}
Lai D., 1997, ApJ, 490, 847 

\bibitem[Lee \& Murakami]{LM19}
Lee U., Murakami D., MNRAS, 2019, 488, 1960 




\bibitem[Ogilvie]{O14}
Ogilvie,  G.I., 2014, Annu. Rev. Astron. Astrophys., 52, 171

\bibitem[Ogilvie \& Lin]{OL04}
Ogilvie, G. I., Lin D. N. C., 2004, ApJ, 610, 477

\bibitem[Ogilvie \& Lin]{OL04}
Ogilvie, G. I., Lin D. N. C., 2007, ApJ, 661, 1180


\bibitem[Press et al.]{NR92}
Press W. H., Teukolsky S. A., Vetterling W. T., Flannery B. P., 1992, 
Numerical Recipes in FORTRAN. Cambridge Univ. Press, Cambridge

\bibitem[Savonije \& Papaloizou]{SP84}
Savonije G.J., Papaloizou J.C.B., 1984, MNRAS, 207, 685

\bibitem[Showman \& Polvani]{SP11}
Showman, A. P., Polvani, L. M., 2011, ApJ, 738, 71

\bibitem[Thorngren et al. ]{TGF19}
Thorngren D., Gao, P., Fortney, J. J.,  ApJL, 2019, 884, 6

\bibitem[Unno et al.]{Unno89}
Unno W., Osaki Y., Ando H., Saio H., Shibahashi H., 1989, Nonradial Oscillations of Stars, 2nd ed., University of Tokyo Press, Tokyo

\bibitem[Watkins \& Cho]{WC10}
Watkins C., Cho J. Y. K., 2010, ApJ, 714, 904

\bibitem[Yang et al.]{Yang19}
Yang J., et al., 2019, ApJ, 575, 46

\bibitem[Youdin \& Mitchell]{YM10}
Youdin, A.,  Mitchell, J. 2010, ApJ, 721, 1113

\bibitem[Yu]{Yu17}
Yu, C., 2017, ApJ, 850, 198

\bibitem[Zahn ]{Z75}
Zahn J. P., 1975, A\&A, 41, 329

\bibitem[Zahn ]{Z77}
Zahn J. P., 1977, A\&A, 57, 383

\bibitem[Zhang et al. ]{Z15}
Zhang K. K., Kong D. L., \& Schubert, G., 2015, ApJ, 806, 270




%
%


\end{thebibliography}

\end{document}